\documentstyle[12pt,epsf]{article}

\def\be{\begin{eqnarray}}
\def\ee{\end{eqnarray}}

\def\He#1{{}$^{#1}${\rm He}}
\def\eps{\varepsilon}

\def\hw{\hbar\omega}

\def\k{{\bf k}}
\def\q{{\bf q}}
\def\p{{\bf p}}

\def\drho{\partial}

\def\grr{G_{\rho\rho}}
\def\gdr{G_{\drho\rho}}
\def\grd{G_{\rho\drho}}
\def\gdd{G_{\drho\drho}}

\def\grrd{G_{\rho;\rho\drho}}
\def\gdrd{G_{\drho;\rho\drho}}

\def\Gleft{\big\langle\!\big\langle}
\def\Gright{\big\rangle\!\big\rangle}

\def\opA{\,\,\hat{\!\!\cal{{A}}\,}}

\begin{document}

\title{Effective Mass of Atom and Excitation Spectrum in Liquid Helium-4 at $T=0$~K}
\author{A.~A.~Rovenchak\\
{\it Department for Theoretical Physics, Ivan Franko National University of Lviv}\\
12 Draghomanov St., Lviv, UA-79005, Ukraine\\
tel: +380 322 979443; e-mail: {\tt andrij@ktf.franko.lviv.ua}
}
\maketitle

\begin{abstract}
A self-consistent approach is applied for the calculations within
the two-time temperature Green functions formalism in the random phase approximation.
The effective mass of \He4 atom is computed as $m^*=1.58\,m$.
The excitation spectrum is found to be in a satisfactory agreement with the
experiment.
The sound velocity is calculated as 230~m/s. The temperature of the Bose-condensation with the
effective mass taken into consideration is estimated as 1.99~K.


\end{abstract}

\section{Introduction}
The idea of the effective mass of helium atom was suggested by Feynman in 1953~\cite{Feynman53}.
He stated that one should insert the
effective mass (being slightly larger than the mass of a `pure' atom $m$) in the
expressions for the density matrix.

Isihara and Samulski~\cite{Isihara77} used the effective mass $m^*=1.71\,m$ in
order to obtain a good agreement of the excitation spectrum sound branch
with the experimental data on the sound velocity.

While in both these cases the effective mass was introduced
phenomenologically,
it appeared to be possible to receive the value of this quantity
basing on the experimental data for the structure factor of
liquid helium-4~\cite{Visn}, the result of $1.70\,m$ was calculated
there by Vakarchuk.

In the paper presented we will give the way of the receiving
\He4 atom effective mass by means of a self-consistent equation.
The expressions are written within the collective variables
formalism as described by Bogoliubov and Zubarev~\cite{BogZub55}.
Two-time temperature Green functions~\cite{Zubarev} are utilized for the
calculation of the thermodynamic averages.

The main idea of the paper is to show the possibility of an essentially simple
approach to the problem of the many-boson system with strong
interaction, such as liquid helium-4. The method applied does not
require much computational efforts. This advantage allows for the
development of the further approximations.

In addition, if one accepts the assumption that the phenomena in
liquid \He4 are at least partly due to the Bose-condensation being
`spoiled' by the interatomic interaction, it turns to be possible
to estimate the lambda transition temperature as the critical
temperature of the ideal Bose-gas. We show that such an approach
leads to a very good agreement with the experiment: if the effective mass is about 50\%
larger than the pure one the Bose-condensation temperature decreases to the value of $\simeq2$~K.

The Green functions technique provides also a possibility to
receive the excitation spectrum of the system.
As a result of the mass renormalization, the excitation spectrum
is found to be in the better agreement with the
experiment in comparison with the Bogoliubov's or Feynman's
one while the expressions are the same (for the latter two spectra
the problem of the so called `roton' minimum overestimation
is well-known if one considers the pure mass).

A self-consistent approach was recently applied for the
calculation of the \He4 excitation spectrum by Pashitskii {\it et
al}~\cite{Pash2002}. The results of this paper are in an excellent
agreement with the experiment. The authors used the `semitransparent
spheres' potential for the calculation with some adjusting
parameters. In our work, we utilize the potential~\cite{MyPotential}
as an input information for the computations. This potential was
received on the basis of the quantum-mechanical equations with the
static structure factor as the only experimental data. Since this
quantity is quite easily measured directly in the scattering
experiments, we consider it as a good approach. Unfortunately, a
direct calculation of the potential for the many-body problem
cannot be made for the time being.

The paper is organized as follows. The calculating procedure is
given in Section~2. The Hamiltonian is written and the equations
of motion for the Green functions are solved in random phase
approximation (RPA) providing a self-consistent equation for the
effective mass extraction. The numerical results are adduced
in Section~3 together with the discussion.

\section{Calculation procedure}

The Hamiltonian of the Bose-system in the collective variables representation reads~\cite{BogZub55}:
\be \label{H-def}
\hat H=\sum_{\k\neq0}
 \left\{\eps_k\Big[\rho_\k\drho_{-\k}-\drho_\k\drho_{-\k}\Big]
  +{N\over2V}\nu_k \Big[\rho_\k\rho_{-\k}-1\Big]
 \right\}
+{1\over\sqrt{N}}\mathop{\sum_{\k\neq0}\sum_{\q\neq0}}\limits_{\k+\q\neq0}
 {\hbar^2\over2m}\k\q\,
 \rho_{\k+\q}\drho_{-\k}\drho_{-\q}
,
\ee
where the operator $\drho_\k=\partial/\partial\rho_{-\k}$. Here,
$\eps_k$ is the energy spectrum of a free particle,
$\eps_k=\hbar^2k^2/2m$, $N$ is the total number of particles in
the system, and $V$ is the system volume. In the thermodynamic
limit, $N/V=\varrho=\rm const$. $\nu_k$ is the Fourier transform
of the interatomic potential.
The item with one summation over the wave vector $\k$ in Eq.~(\ref{H-def}) corresponds to the random
phase approximation, and the second one is the correction.
Let us assume that our system is described by exactly RPA
Hamiltonian $\hat H^{(*)}$, i.~e.,
\be \label{H_eff-def}
\hat H^{(*)}=
\sum_{\k\neq0}
 \left\{\eps_k^*\Big[\rho_\k\drho_{-\k}-\drho_\k\drho_{-\k}\Big]
  +{N\over2V}\nu_k \Big[\rho_\k\rho_{-\k}-1\Big]
 \right\}
\ee
where $\eps_k^*={\hbar^2k^2/2m^*}$
and $m^*$ is the effective mass of \He4 atom. It is the only
quantity suitable for the `effective' role since we wish to
preserve the interatomic potential as the initial information.

One can define $m^*$ demanding that the effective Hamiltonian
(\ref{H_eff-def}) leads to the same ground-state energy as the
initial Hamiltonian (\ref{H-def}), $\langle \hat H^{(*)}\rangle=\langle\hat
H\rangle$:

\be \label{E=E}
&&\!\!\!\!\!\!\!\!\!\!\!\!\sum_{\k\neq0}
 \left\{\eps_k^*\Big[\langle\rho_\k\drho_{-\k}\rangle^{(*)}-\langle\drho_\k\drho_{-\k}\rangle^{(*)}\Big]
  +{N\over2V}\nu_k \Big[\langle\rho_\k\rho_{-\k}\rangle^{(*)}-1\Big]
 \right\}=
 \nonumber\\
&&\!\!\!\!\!\!\!\!\!\!\!\!=\sum_{\k\neq0}
 \left\{\eps_k\Big[\langle\rho_\k\drho_{-\k}\rangle-\langle\drho_\k\drho_{-\k}\rangle\Big]
  +{N\over2V}\nu_k \Big[\langle\rho_\k\rho_{-\k}\rangle-1\Big]
 \right\}
+{1\over\sqrt{N}}\mathop{\sum_{\k\neq0}\sum_{\q\neq0}}\limits_{\k+\q\neq0}
 {\hbar^2\over2m}\k\q\,
 \langle\rho_{\k+\q}\drho_{-\k}\drho_{-\q}\rangle
,
\ee
where the superscript $(*)$ near the angle brackets is introduced
for the convenience.

One can find the operators product average utilizing two-time
temperature Green functions defined as follows~\cite{Zubarev}:
\be
\Gleft A(t)|B(t')\Gright=i\theta(t-t')\langle[A(t),B(t')]\rangle
\ee
with operators given in the Heisenberg representation, $\theta$ is
the Heaviside step function.

\be
\langle AB\rangle\equiv\opA G_{BA}(\hbar\omega)={i\over\hbar}\int_{-\infty}^{+\infty}d\hbar\omega
{G_{BA}(\hbar\omega+i\eps)-G_{BA}(\hbar\omega-i\eps)\over
e^{\beta\hbar\omega}-1}\Biggl|_{\eps\to+0}
\ee
where $G_{BA}$ stands for $\Gleft B|A \Gright$ and the operator
$\opA$ is introduced for the convenience. We put the time arguments in
the operators $A(t), B(t')$ to coincide: $t-t'=0$. This will
provide static properties of the system under consideration. In
the above expression, $\beta$ is the inverse temperature,
$\beta=1/T$.

Now, we will
proceed to the equations of motion for the Green functions
$\grr(\k) \equiv \Gleft \rho_{\k} | \rho_{-\k} \Gright, \newline
 \grd(\k) \equiv \Gleft \rho_{\k} | \drho_{-\k} \Gright$, etc.
It is easy to receive the following set of equations in RPA:
\be \label{Gk}
&&(\hw+\eps_k)\grr(\k)=2\eps_k\gdr(\k),  \nonumber \\
&&(\hw-\eps_k)\gdr(\k)=\varrho\nu_k\grr(\k)+{1\over 2\pi}, \nonumber\\
&&(\hw+\eps_k)\grd(\k)=2\eps_k\gdd(\k)-{1\over 2\pi}, \nonumber\\
&&(\hw-\eps_k)\gdd(\k)=\varrho\nu_k\grd(\k).
\ee

Next, let us consider the triple product average $\langle ABC\rangle$.
One can obtain it utilizing either the Green function
$G_{C;AB}\equiv\Gleft C|AB\Gright$ or
$G_{BC;A}\equiv\Gleft BC|A\Gright$. We suggest the first
possibility to fulfill:
\be
\langle ABC\rangle=\opA G_{C;AB}\equiv\Gleft C|AB\Gright.
\ee
In other words, we neglect the functions of the type
$G_{BC;A}\equiv\Gleft BC|A\Gright$ for the sake of simplicity
(when applying this to Eq.~(\ref{Gk}) it means
that only the RPA term of the Hamiltonian (\ref{H-def}) is taken into consideration
when constructing the equations of motion).

Having performed the similar procedure with the function
$G_{\drho;\rho\drho}(\k_1,\k_2,\k_3)\equiv\Gleft \drho_{\k_1}|\rho_{\k_2}\drho_{\k_3}\Gright$
we obtain in RPA the following set of equations:
\be \label{G123}
&&(\hw-\eps_{k_2})\gdrd(-\k_2,\k_1+\k_2,-\k_1)=\varrho\nu_{k_2}\grrd(-\k_2,\k_1+\k_2,-\k_1)
                                              -g_{\drho\rho\drho}(\omega)\nonumber\\
&&(\hw+\eps_{k_2})\grrd(-\k_2,\k_1+\k_2,-\k_1)=2\eps_{k_2}\gdrd(-\k_2,\k_1+\k_2,-\k_1)
                                              -g_{\rho\rho\drho}(\omega),
\ee
where the quadruple Green functions were decoupled in such a way providing for
the inhomogeneous set of equations:
\be
\Gleft AB | CD\Gright=
          \langle BD\rangle \Gleft A|C \Gright
         +\langle CA\rangle \Gleft B|D \Gright
         +\langle AD\rangle \Gleft B|C \Gright
         +\langle CB\rangle \Gleft A|D \Gright.
\ee
The inhomogeneous terms in Eq.~(\ref{G123}) read:
\be
g_{\rho\rho\drho}(\omega)&=&{\hbar^2\over2m}\Bigg[
  \k_1\k_2 \, 2D_{k_1}\grr(\k_1+\k_2) + \k_1\k_2 \, S_{|\k_1+\k_2|}\gdd(\k_1)\nonumber\\
 &&\quad{}+\k_2(\k_1+\k_2)\, D''_{k_1}\gdr(\k_1+\k_2)+\k_2(\k_1+\k_2)\, D''_{|\k_1+\k_2|}\grd(\k_1)
  \Bigg]                  \nonumber\\
g_{\drho\rho\drho}(\omega)&=&{\hbar^2\over2m}\Bigg[
  \k_1(\k_1+\k_2)\, D_{k_1}\gdr(\k_1+\k_2) +\k_1(\k_1+\k_2)\,
  D''_{|\k_1+\k_2|}\gdd(\k_1)\Bigg]\times2.
\ee
The notations for the averages of pair products are listed below:
\be  \label{pairAverages}
\langle\rho_{-\k}\rho_{\k}\rangle&\equiv& S_k = {1\over\alpha_k}\coth{\eps_k\alpha_k\over2T},\nonumber\\
\langle\rho_{-\k}\drho_{\k}\rangle &\equiv& D''_k = {1\over2}\left({1\over\alpha_k}\coth{\eps_k\alpha_k\over2T}-1\right)
={1\over2}\left(S_k-1\right),\nonumber\\
\langle\drho_{-\k}\drho_{\k}\rangle&\equiv& D_k = {1-\alpha_k^2\over4\alpha_k}\coth{\eps_k\alpha_k\over2T}
=-{\varrho\nu_k\over2\eps_k}\,S_k.
\ee
The quantity $\alpha_k$ is defined as follows:
\be
\alpha_k=\left(1+2\varrho\nu_k/\eps_k\right)^{1/2}.
\ee

Now, if we turn back to correlation (\ref{E=E}), the meaning of
the asterisk as a superscript becomes clear: one should substitute
$m$ with $m^*$ in the l.h.s. of this equation.

In the ground state ($T=0$~K), hyperbolic cotangents in Eq.~(\ref{pairAverages}) equal to
unity.
Therefore, a self-consistent equation for the extraction of $m^*$ becomes as follows:
\be \label{self-consistent}
&&-{1\over N}\sum_{\k\neq0} {\eps_k^*\over4}\left(\alpha_k^*-1\right)^2=
-{1\over N}\sum_{\k\neq0} {\eps_k\over4}\left(\alpha_k-1\right)^2 \nonumber\\
&&-
{1\over N^2}\mathop{\sum_{\k\neq0}\sum_{\q\neq0}}\limits_{\k+\q\neq0}
 \left(\hbar^2\over2m\right)^2{\k\q\over\alpha_\k\alpha_q\alpha_p}
 \Bigg[
  \left({1\over \eps_k\alpha_k+\eps_q\alpha_q} + {1\over \eps_q\alpha_q+\eps_p\alpha_p}\right)
  \left(\k\q{\varrho\nu_k\varrho\nu_q\over\eps_k\eps_q}
       +\k\p{\varrho\nu_k\over2\eps_k}
       +\q\p{\varrho\nu_q\over2\eps_q}
  \right) \nonumber\\
 &&{}+
 \k\p{\varrho\nu_k\over2\eps_k}
 \left({-\alpha_p\over\eps_k\alpha_k+\eps_q\alpha_q}+{\alpha_q\alpha_p\over\eps_q\alpha_q+\eps_p\alpha_p}\right)
 -\q\p{\varrho\nu_q\over2\eps_q}
 \left({\alpha_p\over\eps_k\alpha_k+\eps_q\alpha_q}+{\alpha_k\over\eps_q\alpha_q+\eps_p\alpha_p}\right)
 \Bigg]
\ee
where $\p=\k+\q$. We also consider the specific energy
instead of the total one
by introducing the factor of $1/N$.

\section{Numerical results and discussion}
We use the previously obtained results \cite{MyPotential} for the
interatomic potential Fourier transform $\nu_k$. The value of the
equilibrium density is $\varrho=0.02185$~\AA$^{-3}$. The mass of
helium-4 atom equals $m=4.0026$~a.m.u.
We transit from the summation over the wave vector to the integration in the usual way:
$\sum_\k \to V\int d\k/(2\pi)^3$.
The value of the upper cut-off for the
integration over the wave vector is 16.0~\AA$^{-1}$.

The solution of Eq.~(\ref{self-consistent}) at the abovelisted conditions is
\be
m^*=1.58\,m.
\ee

\begin{figure}[h]
\epsfxsize=100mm
\centerline{\epsfbox{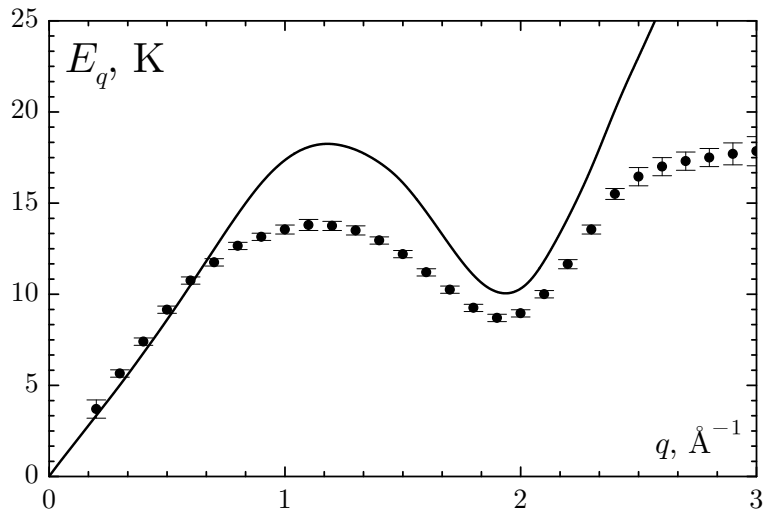}}
\bigskip
\caption{ \label{EnEff}
Excitations spectrum of liquid helium-4.
Filled circles --- experimental data~\cite{ExpEnergy};
Solid line --- calculated energy.
}
\end{figure}

One can also receive the excitations spectrum using Green
functions. The solutions of set (\ref{Gk}) are proportional
to $1/(\hbar{}^2\omega{}^2-\eps_k^2\alpha_k^2)$ providing the spectrum
$E_k=\pm\eps_k\alpha_k$ --- a very well known result~\cite{BogZub55}. If one inserts
the effective mass into the definitions of $\eps_k$ and $\alpha_k$
the obtained curve fits the experimental one quite satisfactory,
see Fig.~\ref{EnEff}.

\bigskip
\bigskip

The phonon branch is reflected quite good providing the sound
velocity of approximately 230~m/s {\it vs} the experimental
238~m/s at $T=0.8$~K~\cite{NIST} or 240~m/s at $T=0.1$~K~\cite{NBS}.
The so called `roton' minimum also has the value close to the
experimental one.

In addition, the obtained value of the effective
mass shifts the temperature of the Bose-condensation from $T_c=3.14$~K for the pure mass
to $T_c=1.99$~K {\it vs} the experimental temperature of the lambda
transition $T_\lambda=2.17$~K. We consider the results discussed
above as quite good as for such a rough approximation as random
phases.

\section*{Acknowledgements}
The author is grateful to Prof. I.~Vakarchuk for the valuable
discussions on the problem considered in this work.


\begin{thebibliography}{99}
\bibitem{Feynman53}R.~P.~Feynman, Phys. Rev. {\bf 91}, 1291 (1953).
\bibitem{Isihara77}A.~Isihara and T.~Samulski, Phys.~Rev. B {\bf 16},
1969 (1977).
\bibitem{Visn}I.~O.~Vakarchuk, Visn. Lviv. univer. Ser. fiz. {\bf 26}, 29 (1993) [in Ukrainian].
\bibitem{BogZub55}N.~N.~Bogoliubov and D.~N.~Zubarev, Zhurn.~Eksp.
Teor.~Fiz. {\bf 28}, 129 (1955) [Sov Phys.--JETP {\bf 1}, 83
(1955)].
\bibitem{Zubarev}D.~N.~Zubarev, {\it Neravnovesnaia statisticheskaia
termodinamika} (Nauka, Moscow, 1971) [in Russian];
D.~N.~Zubarev, {\it Nonequilibrium statistical thermodynamics} (Consultants Bureau, New York,
1974).
\bibitem{Pash2002}E.~A.~Pashitskii, S.~I.~Vilchinskyy,
S.~V.~Mashkevich, Fiz.~Nizk. Temp. {\bf 28}, 115 (2002) [in Russian].
\bibitem{MyPotential}I.~O.~Vakarchuk, V.~V.~Babin, A.~A.~Rovenchak,
J. Phys. Stud. (Lviv) {\bf 4}, 16 (2000).
\bibitem{ExpEnergy}R.~A.~Cowley and A.~D.~B.~Woods, Can. J. Phys. {\bf 49} 177 (1971).
\bibitem{NIST}V.~D.~Arp, R.~D.~McCarty, D.~G.~Friend,
Natl. Inst. Stand. Technol. Tech. Note 1334 (revised) (1998).
\bibitem{NBS}R.~D.~McCarty, Natl. Bur. Stand. Tech. Note 1029 (1980).
\end{thebibliography}
\end{document}